\documentclass{raa}
\usepackage{graphicx,times}
\usepackage{natbib}
\usepackage{amssymb,amsmath}
\bibpunct{(}{)}{;}{a}{}{,}

\usepackage[a4paper=true,dvipdfm=true,pagebackref=true]{hyperref}
\hypersetup{pdftitle = The title of my PDF, pdfauthor = My name, pdfsubject= The subject, pdfkeywords = keyword1 keyword2 keyword3}
\hypersetup{colorlinks = true, linkcolor = green, anchorcolor = red, citecolor = blue, filecolor = red, pagecolor = red, urlcolor = red}
\makeatletter
\newcommand*{\rom}[1]{\expandafter\@slowromancap\romannumeral #1@}
\makeatother

\def\fm{\hbox{$.\!\!^m$}}
\def\fs{\hbox{$.\!\!^s$}}
\def\degr{\hbox{$^\circ$}}

\begin{document}%

\title{Physical and Geometrical Parameters of CVBS~\rom{11}: COU1511 (HIP12552)}

\volnopage{ {\bf 2016} Vol.\ {\bf X} No. {\bf XX}, 000--000}
   \setcounter{page}{1}

   \author{ Mashhoor A. Al-Wardat\inst{1}, M. H. El-Mahameed \inst{2}, Nihad A. Yusuf \inst{3,4},  Awni M. Khasawneh \inst{5}, Suhail G. Masda\inst{6}}

\institute{  Physics Department, Al al-Bayt University, PO Box: 130040, Mafraq, 25113 Jordan; {\it mwardat@aabu.edu.jo}   \\
         \and	Department of Physics, Al-Hussein Bin Talal University, Maan, 71111 Jordan
    	 \and    Department of Physics, Yarmouk University, Irbid, 21163,  Jordan
         \and Princess Sumaya University for Technology, Amman, Jordan
         \and    Royal Jordanian Geographic Center, Amman, 11941 Jordan  \\       	
         \and    Astrophysikalisches Institut und Universit\"{a}ts-Sternwarte Jena, FSU Jena, 07745 Jena, Germany\\
\vs \no
   {\small Received 2016 May 4; accepted 2016 June 8}
}

\abstract{
Model atmospheres of the close visual  binary star COU1511 (HIP12552) are constructed using  grids of Kuruz's blanketed models to build the individual synthetic SEDs for both components. These synthetic SED's are combined together for the entire system and compared with the observational one following Al-Wardat's complex method for analyzing close visual binary stars. The entire observational spectral energy distribution (SED) of the system is used as a reference for comparison between synthetic SED and the observed one. The parameters of both components  are derived as: $T_{\rm eff}^{a}
=6180\pm50 $\,K, $T_{\rm eff}^{b} =5865\pm70$\,K,  log $g_{a}=4.35\pm0.12$,
log $g_{b}=4.45\pm0.14$, $R_{a}=1.262\pm0.08R_\odot$, $R_{b}=1.006\pm0.07R_\odot$,  $L_a=2.09\pm0.10 L_\odot$, $L_b=1.08\pm0.12L_\odot$, with spectral types F8 and G1 for both components (a,b) respectively, and age of $3.0\pm 0.9$ Gy. A modified orbit of the system is built and the masses of the two components are calculated as $M_a=1.17\pm0.11M_\odot$, $M_b=1.06\pm0.10M_\odot$.
}
%%\pacs{95.75.Fg, 97.10.Ex, 97.10.Pg, 97.10.Ri, 97.20.Jg, 97.80.Fk}

\authorrunning{Al-Wardat et al. }            %author_head in even pages
   \titlerunning{Parameters of CVBS ~\rom{11}: COU1511 (HIP12552)}  % title_head in odd pages
\maketitle%

\section{INTRODUCTION }

 Recent surveys of the sky showed that more than 50\% of the galactic stellar systems are binaries, which raises their importance in understanding the formation and evolution of the galaxy. This role in determining precisely different stellar parameters gives the study of binary stars a special importance. The case is a bit complicated in the case of close visual binary stars (CVBS), which are not resolved as binaries by inspecting limited images, but can be resolved in space based observations, or by using modern techniques of ground-based observations, like speckle interferometry and adaptive optics.

 In addition to that, \cite{2001A&A...366..868D} pointed out that the study of orbital motion of visual and interferometric pairs remains an important astronomical discipline. The visual binaries are the main key source of information about stellar masses and distances, and they define practically our understanding of stellar physical properties especially for the lower part of the main sequence stars .

 For now, hundreds of CVBS with periods on the order of 10 years or less are routinely observed by different groups of high resolution techniques around the world. This has helped in determining the orbital parameters and magnitude differences for some of these CVBS. However, this is not sufficient to determine the individual physical parameters of the components of the system.

 Al-Wardat's method for analyzing CVBS \citep{2012PASA...29..523A} offers a complementary solution for this problem by implementing differential photometry, spectrophotometry, atmospheric modeling, and orbital solution in  accurate determination of different physical and geometrical parameters of this category of stars. The method was successfully applied to several solar type and subgaint binary systems such as Hip70973, Hip72479 \citep{2012PASA...29..523A}, Hip689 \citep{2014AstBu..69...58A}, and Hip11253 \citep{2009AstBu..64..365A}. %The selected system must fulfill the following requirements of the method:
 %\begin{itemize}
 %	\item Precise magnitude difference measurements.
 %	\item Observational SED must cover the optical range.
 %	\item Photometrical magnitude measurements.
 %\end{itemize}

 As a consequence of the previous work, this paper (the XI in its series)  presents the  analysis of the nearby solar type CVBS COU1511 (Hip12552), with a modification to its parallax.

 %using Al-Wardat's method which is based on combining magnitude difference measurements of speckle interferometry,  entire spectral energy distribution (SED) of spectrophotometry, all along with atmospheres modeling to estimate the individual physical parameters.

 %which is a well-known interferometric binary.

 Table~\ref{tabl1} shows the basic data of the system taken from SIMBAD and NACA/IPAC catalogues, and Table~\ref{tabl2} shows data from Hipparcos and Tycho catalogues~\citep{1997yCat.1239....0E}, while  Table~\ref{tabl4} shows the magnitude difference of the system along with filters used to observe them.

 %Table~\ref{tabl3} shows the  orbital elements of the system Hip12552 taken from the sixth interferometric catalogue http://www.us no.navy.mil/U SNO/astrometry/optical-IR-prod/wds/orb6,

 \begin{table}[!h]
 	\begin{center}
 		\caption{Basic data of the system} \label{tabl1}
 		\begin{tabular}{lcc}\hline\hline
 			& Hip12552 & Reference
 			\\
 			\hline
 			$\alpha_{2000}$  & $02^h 41^m 28\fs88$&SIMBAD$^\dagger$\\
 			$\delta_{2000}$ &$+40\degr52' 50.''84$&-\\
 			Tyc.& 2849-1282-1 & -\\
 			HD & 16656 & -\\
 			Sp. Typ. & G0&-\\
 			$E(B-V)$ & $0.076\pm 0.002$& NASA/IPAC$^{*}$\\
 			$A_v$&$0\fm24$& NASA/IPAC
 			\\
 			\hline\hline
 		\end{tabular}
 		\\
 		$^\dagger$ http://simbad.u-strasbg.fr/simbad/\\
 		%$^\ddagger$ http://irsa.ipac.caltech.edu/frontpage/
 		$^{*}$~http://irsa.ipac.caltech.edu,
 	\end{center}
 \end{table}

 \begin{table}[!h]
 	\begin{center}
 		\caption{Data of Hipparcos and Tycho Catalogues} \label{tabl2}
 		\begin{tabular}{lcc}\hline\hline
 			& Hip12552 & Source of data
 			\\
 			\hline
 			$V_J(Hip)$ & $8\fm51$& Hipparcos
 			\\
 			$(B-V)_J(Hip)$ & $0\fm60\pm0.018$& -
 			\\
 			$\pi_{Hip}$ (mas) & $9.69\pm1.29$ & -\\
 			$B_T$ & $9\fm24\pm0.016$ & Tycho
 			\\
 			$V_T$ & $8\fm59\pm0.014$& -\\
 			$\pi_{Tyc}$ (mas) & $14.7\pm9.20$& -\\
 			$\pi_{Hip}$$^{*}$ (mas) & $11.07\pm1.07$ & New Hipparcos\\
 			\hline\hline
 		\end{tabular}
 		\\
 		%$^{*}$~http://irsa.ipac.caltech.edu,
 		$^{*}$~\citep{2007A&A...474..653V}
 	\end{center}
 \end{table}

 \begin{table}[!ht]
 	\begin{center}
 		\caption{Magnitude difference between the components of the system Hip12552, along with the  filters used to obtain the observations. }
 		\label{tabl4}
 		\begin{tabular}{lcc}
 			\noalign{\smallskip}
 			\hline\hline
 			%	\noalign{\smallskip}
 			$\triangle m $& Filter ($\lambda/\Delta\lambda$)& References  \\
 			\hline
 			%	\noalign{\smallskip}
 			$0\fm65\pm 0.06$ &  $545nm/30 $& 1   \\						
 			$0\fm75\pm0.04$ &  $545nm/30 $&  2   \\		
 			$0\fm88\pm-$&  $562nm/40 $&  3    \\
 			$0\fm86\pm-$ &  $692nm/40$ &   3  \\		
 			\hline\hline
 		\end{tabular}
 		\\
 		$^1${\cite{2007AstBu..62..339B}},
 		$^2${\cite{2006AJ....132..994D}},
 		$^3${\cite{2011AJ....141...45H}}.
 	\end{center}
 \end{table}
%\section{Analysis and Method of the system}
\section{Atmospheric Modeling}
\label{Atmospheric Modeling}
The observational SED of the system Hip12552 obtained by~\cite{2002BSAO...53...58A} is used as reference for the comparison with synthetic SED.

\noindent Using $\ m_V=8\fm51$ (see Table~\ref{tabl2}), $\Delta m $ $=0\fm76\pm 0.03$ (as the average of all $\Delta m $ using the different filters for V-band only (545-562 nm), see Table~\ref{tabl4}), and Hipparcos trigonometric parallax ($\pi=11.83\pm 1.07$ mas), the individual and absolute magnitudes of both components (a, b) of the system are calculated using the following relations:
{\begin{eqnarray}
	\label{eq1}
	\frac{F_a}{F_b}=2.512^{-\triangle m}
	\end{eqnarray}}
\begin{eqnarray}
\label{eq3}
\ M_v=m_v+5-5\log(d)-A_v
\end{eqnarray}
to get $m_v^a=8\fm95\pm0.02$, $m_v^b=9\fm71\pm0.05$, %from which the individual absolute magnitudes are calculated yielding
and $ M_v^a=4\fm07\pm0.18$, $ M_v^b=4\fm83\pm0.19$, where the extinction value $ A_v$ was taken from Table~\ref{tabl1}.

To calculate the preliminary input parameters used to build  the atmospheric modelling, we use the bolometric magnitudes, the luminosities from~\cite{1992adps.book.....L}, and ~\cite{2005oasp.book.....G} with the following relations:

\begin{eqnarray}
\label{eq8}
\log(R/R_\odot)= 0.5 \log(L/L_\odot)-2\log(T_{eff}/T_\odot),
\end{eqnarray}
%getting the values $ R_{a}=1.40R_\odot$ and $R_{b}=1.2R_\odot ,$. These values of radii with estimated masses enable us to obtain the gravity acceleration at the surface of the components using the following relation:
\begin{eqnarray}
\label{eq5}
\log g = \log(M/M_\odot)- 2\log(R/R_\odot) + 4.43,
\end{eqnarray}
to estimate the effective temperatures and gravity acceleration. These  values for the effective temperature and gravity acceleration allow us to construct  the model atmosphere for each component using grids of Kuruz's line blanketed models (ATLAS9)~\citep{1994KurCD..19.....K}. Here we used $T_\odot=5777\rm{K}$ and $M_{bol}^\odot=4\fm75$ in all calculations.

The total energy flux from a binary star is due to the net luminosity of the components a, and b located at a distance d from the Earth. The total energy flux may be written as:
\begin{eqnarray}
\label{eq33}
F_\lambda \cdot d^2 = H_\lambda ^a \cdot R_{a} ^2 + H_\lambda ^b
\cdot R_{b} ^2,
\end{eqnarray}
\noindent Rearranging equ.~\ref{eq33} gives
\begin{eqnarray}
\label{eq44}
F_\lambda  = (R_{a} /d)^2(H_\lambda ^a + H_\lambda ^b \cdot(R_{b}/R_{a})^2) ,
\end{eqnarray}
where $ R_{a}$ and $ R_{b}$ are the radii of the primary and secondary components of the system in solar units, $H_\lambda ^a $ and  $H_\lambda ^b$ are the fluxes at the surface of the star and $F_\lambda$ is the flux for the entire SED of the system.

%From preliminary calculated parameters, it is observed that the maximum values of the synthetic SED in the left part of the SED do not coincide with the observed one. This means that the effective temperature of both components of the system, radii, and/or the parallax of the system should be smaller.

\begin{figure}[!ht]
	\centering
	% 		 \centerline{\psfig{figure=HIP95995all.eps,width=0.5\textwidth,clip=}}
	\includegraphics[angle=0,width=14cm]{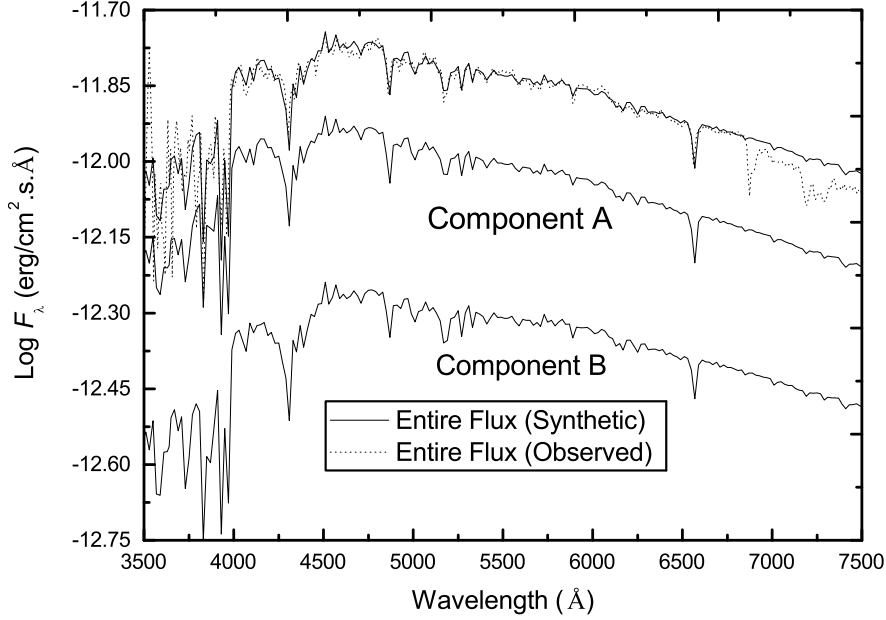}
	\caption{Best fit between the entire observed spectrum (dotted line) ~\citep{2002BSAO...53...58A} and the synthetic entire SED (solid line) for the system Hip12552 using  the following parameters $T_{\rm eff}^A =6180\pm50$\,K, log $g_A=4.35\pm0.10, R_A=1.262\pm0.09R_\odot$, $T_{\rm eff}^B =5865\pm70$\,K, log $g_B=4.45\pm0.15, R_B=1.006\pm0.10 R_\odot $  with $d=84.53\pm0.009$\ pc ($\pi=11.83 $ mas).} \label{fig.}
\end{figure}

Many attempts were made to achieve the best fit (Fig.~\ref{fig.}) between the observed flux of~\cite{2002BSAO...53...58A} and the total computed one using the iteration method of different sets of parameters. The best fit is found using the following set of parameters:
$$ T_{\rm eff}^{a}
=6180\pm50{\rm K}, T_{\rm eff}^{b} =5865\pm70{\rm K},$$
$$ \log g_{a}=4.35\pm 0.12,  \log g_{b}=4.45\pm 0.14,$$
$$R_{a}=1.262\pm0.08R_\odot, R_{b}=1.006\pm0.07R_\odot$$
%$d=96.0$\,pc ($\pi=10.42\pm0.2 $ mas).
%This gives using equ.~\ref{eq5}: $M_a=1.46\pm0.30M_\odot$  and $ M_b=1.23\pm0.28 M_\odot$, and
Using equ.~\ref{eq8} the luminosities  are calculated yielding the following values: $L_a=2.09\pm0.10L_\odot$ and  $L_b=1.08\pm0.12L_\odot$.  Using tables of~\cite{2005oasp.book.....G} or the $Sp-T_{eff.}$ empirical relation from~\cite{1992adps.book.....L} , the spectral types of the components (a, b) of the system are F8 and G2 respectively.% Note that: the best fit can be achieved when changing the radii of the system but this will contradict the radii of tables of~\cite{2005oasp.book.....G} and~\cite{1992adps.book.....L}.

\section{Orbital Analysis}
The orbit of the system is built using the positional measurements listed in Table~\ref{tabl5}, following  Tokovinin's method \citep{1992ASPC...32..573T}. The modified orbital elements of the system along with those taken from the sixth interferometric catalogue are listed in Table~\ref{tabl6}. The table shows a good agreement between our estimated orbital period, P;  eccentricity, e;  semi-major axis, a;  inclination, $ \ i$;  argument of periastron, $ \omega$;  position angle of nodes, $\Omega$; and  time of primary minimum, $\ T_0$ with those previously reported results.

\begin{table}[t]
	\begin{center}
		\centering
		\caption{ Positional measurements of the system from the Fourth Interferometric Catalogue.}
		\small%\centerline
		\label{tabl5}
		\centering
		\begin{tabular}{lrcr}
			\hline
			\multicolumn{1}{c}{Epoch} &
			\multicolumn{1}{c}{$\theta$},deg &
			\multicolumn{1}{c}{$\rho$},arcsec &
			References
			\\
			\hline
			\centering
			1979.7732&	91.6&	0.156& 1\\
			
			1982.7605&	65.9&	0.153&	2\\
			
			1982.7659& 66.3& 0.142&	2\\
			1983.7131&	57.9&	0.136&	2 \\
			1984.7046&	49.4&	0.119&	2 \\
			
			1985.8540&	30.4&	0.106&	2 \\
			
			1991.8973&	184.9*&	0.105&	3 \\
			1993.7652&	161.0*&	0.122&	4\\
			
			1994.7087&	151.3*&	0.136&	5  \\
			1994.8989&	143.0*&	0.146&	6  \\
			
			1995.7710&	139.2*&	0.135&	7   \\
			1996.6912&	132.7*&	0.150&	5  \\
			2000.8730&	98.8&	0.152&	8\\
			2003.9468&	73.8&	0.143&	9 \\
			
			2003.9598&	77.1&	0.146&	10   \\
			
			2004.8374&	65.0&	0.135&	11 \\
			2004.9905&	64.1&	0.135&	12\\
			2007.6075&	30.0&	0.109&	13\\
			2008.861&	334.9*&	0.077&	14\\
			2010.0074&	328.1&	0.0659&	15\\
			\hline
		\end{tabular}
		\smallskip\\
		$^*$ These points were modified by $180\degr$ to achieve consistency with nearby points.\\
		$^1${\cite{1982ApJS...49..267M}},
		$^2${\cite{1987AJ.....93..688M}},
		$^3${\cite{1994AJ....108.2299H}},
		$^4${\cite{1994A&AS..105..503B}},
		$^5${\cite{2000AJ....119.2403T}},
		$^6${\cite{1999A&AS..140..287B}},
		$^7${\cite{1997AJ....114.1639H}},
		$^8${\cite{2006BSAO...59...20B}},
		$^9${\cite{2013AstBu..68...53B}},
		$^{10}${\cite{2008AJ....135.1334H}},
		$^{11}${\cite{2007AstBu..62..339B}},
		$^{12}${\cite{2006AJ....132..994D}},
		$^{13}${\cite{2011AJ....142..176M}},
		$^{14}${\cite{2012AN....333..727G}},
		$^{15}${\cite{2011AJ....141...45H}}.
		
	\end{center}
\end{table}
\section{Masses}
Using the estimated orbital element, the masses of the system and the corresponding errors are calculated using the following relations:
\begin{eqnarray}
	\label{eq31}
	\ M_A +M_B=(\frac{a^3}{\pi^3P^2})\ M_\odot
\end{eqnarray}

\begin{eqnarray}
	\label{eq32}
	\frac{\sigma_M }{M} =\sqrt{(3\frac{\sigma_\pi}{\pi})^2+(3\frac{\sigma_a}{a})^2+(2\frac{\sigma_p}{p})^2}
\end{eqnarray}
The preliminary result obtained using the new Hipparcos trigonometric parallax ($\pi=11.07\pm1.07 $ mas)~\citep{2007A&A...474..653V} is $ \ M_a +M_b $= $2.72\pm0.75M_\odot$, while it is $3.26\pm1.18M_\odot$ when using Hipparcos trigonometric parallax ($\pi=10.42\pm0.2 $ mas, See Table~\ref{tabl6}).
But depending on our analysis (Sec. ~\ref{Atmospheric Modeling}), we achieved the best fit between the synthetic and observational entire SED using  $\pi=11.83$ mas, this new parallax value gives a mass sum of  $\ M_a +M_b $= $2.23\pm0.57M_\odot$, which fits better the positions of the two components on the evolutionary tracks as shown in Fig. ~\ref{evol}.

\begin{table*}[ht]
	\begin{center}
		\caption{Orbital elements of the system.}
		\label{tabl6}
		\begin{tabular}{lccc}
			\noalign{\smallskip}
			\hline\hline
			%\noalign{\smallskip}
			Parameters & This work & \cite{2001IAUDS.145....1H}&\cite{1996IAUDS.128....1C}
			\\
			%check error values
			\hline
			$P$, yr       & $21.90188 \pm 0.07339$   & $22.18$& $19.7$
			\\
			$T_0,$ yr   & $2010.7566 \pm 0.0714$       & $1988.64$& $1988.49$
			\\
			$e$           & $0.4599 \pm 0.0087$   & $0.474$& $0.43$
			\\
			%	$e$           & $0.354 \pm 0.010$   & $0.37 \pm 0.25$& $0.36 \pm 0.145$
			%	\\
			$a, $ arcsec  & $0.1209 \pm0.0017 $           & $0.124 $& $0.121 $
			\\
			$i $, deg           & $ 152.10 \pm 2.71$      & $ 147.0$&  $ 139.7$
			\\
			$\omega$, deg   & $274.90 \pm 6.06$        & $287.0$ & $114.2$
			\\
			$\Omega$, deg & $202.50 \pm 6.32$          & $217.2$ &  $44.50$
			\\
			$  \ M_a +M_b, M_\odot$ &$2.72\pm0.75^{*}$&  &
			\\
			& $2.23\pm0.57^{**}$ &    &
			\\
				& $3.26\pm1.18^{***}$ &    &
				\\
			\hline\hline
			%\noalign{\smallskip}
		\end{tabular}\\
		$^{*}$Based on new Hipparcos trigonometric parallax ($\pi=11.07\pm1.07 $ mas)\\
		$^{**}$Based on the  parallax estimated in this work ($\pi=11.83 $ mas)\\
		$^{***}$Based on Hipparcos trigonometric parallax  ($\pi=10.42\pm0.2 $ mas)\\
	\end{center}
\end{table*}
\begin{figure}[t]
	\centering
	%		 \centerline{\psfig{figure=HIP95995all.eps,width=0.5\textwidth,clip=}}
	\includegraphics[angle=0,width=14cm]{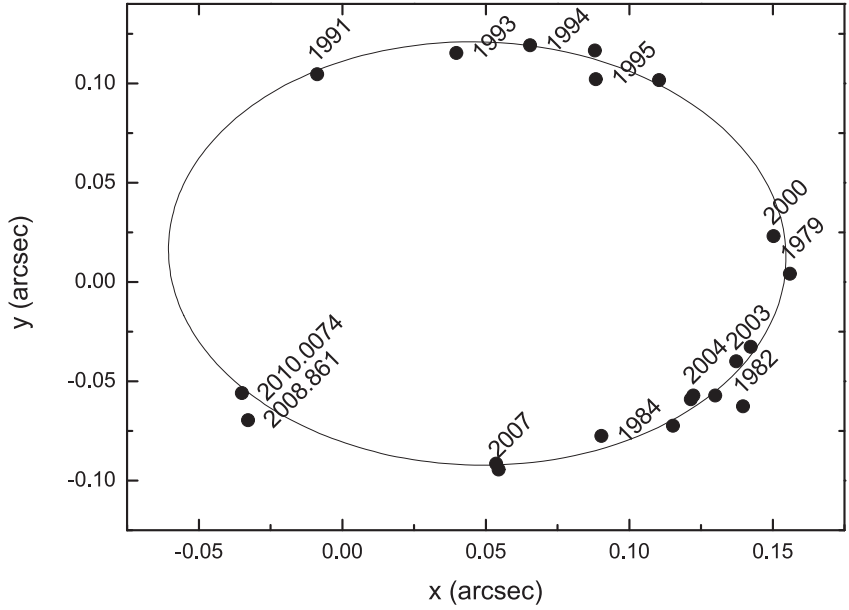}
	\caption{ Relative visual orbit of the system HIP12552 showing the epoch of the positional measurements. } \label{orb22}
\end{figure}
%\begin{figure}[!ht]
%	\centering
%	\includegraphics[angle=0,width=7.5cm]{2.jpg}
%	\caption{ comparison between the modified orbit of this study (solid line) and that taken from sixth interferometric catalogue (dashed line), the solid two points are new points added to the orbit. } \label{orb22}
%\end{figure}
\section{Synthetic Photometry}
As a double-check for the best fit and to present a new synthetic photometrical data of the unseen individual components of the system, we apply the  following relation~\citep{2006AJ....131.1184M, 2007ASPC..364..227M}:
\begin{equation}
	\label{55}
	m_p[F_{\lambda,s}(\lambda)] = -2.5 \log \frac{\int P_{p}(\lambda)F_{\lambda,s}(\lambda)\lambda{\rm d}\lambda}{\int P_{p}(\lambda)F_{\lambda,r}(\lambda)\lambda{\rm d}\lambda}+ {\rm ZP}_p\,,
\end{equation}
to calculate total and individual
synthetic magnitudes of the systems,
where $m_p$ is the synthetic magnitude of the passband $p$, $P_p(\lambda)$ is the dimensionless sensitivity function of the passband $p$, $F_{\lambda,s}(\lambda)$ is the synthetic SED of the object and $F_{\lambda,r}(\lambda)$ is the SED of the reference star (Vega).  Here the zero points (ZP$_p$) of \cite{2007ASPC..364..227M} are adopted.

 Calculated synthetic magnitudes and color indices of the entire system and individual components of different photometrical systems are shown in Table~\ref{synth1}.

%Table 8 shows a comparison between   magnitudes and color indices of the combined system and those taken from [30]. This shows a good consistency within the three systems: Johnson-Cousin, Strӧmgren ,and Tycho.

\begin{table}[!ht]
	\small
	\begin{center}
		\caption{ Synthetic magnitudes and color indices  of the system.}
		\label{synth1}
		\begin{tabular}{lcccc}
			\noalign{\smallskip}
			\hline
			\noalign{\smallskip}
			Sys. & Filter & Entire & Comp.& Comp.\\
			&     & $\sigma=\pm0.03$&   a    &     b      \\
			\hline
			\noalign{\smallskip}
			Joh-  & $U$    & 9.22 & 9.59 & 10.56 \\
			Cou.          & $B$ & 9.11   &  9.51 &  10.38 \\
			& $V$ & 8.51 &  8.95 &  9.71\\
			& $R$ & 8.18 &  8.64 & 9.36  \\
			&$U-B$& 0.11 & 0.08 & 0.18 \\
			&$B-V$& 0.60  &  0.57 &  0.66 \\
			&$V-R$& 0.33&  0.31 & 0.36 \\
			\hline
			\noalign{\smallskip}
			Str\"{o}m. & $u$ & 10.38 & 10.76 &  11.70  \\
			& $v$ & 9.44 & 9.83  & 10.73  \\
			& $b$ & 8.85 & 9.27 &  10.08 \\
			&  $y$& 8.48 & 8.92 & 9.68  \\
			&$u-v$& 0.94 & 0.93 & 0.97 \\
			&$v-b$& 0.59 & 0.56 & 0.65 \\
			&$b-y$& 0.37 & 0.35& 0.40 \\
			\hline
			\noalign{\smallskip}
			Tycho       &$B_T$  & 9.25   & 9.65 & 10.55   \\
			&$V_T$  & 8.58   & 9.01 & 9.79  \\
			&$B_T-V_T$& 0.68 & 0.64 & 0.76\\
			\hline
			\noalign{\smallskip}
		\end{tabular}
	\end{center}
\end{table}

\begin{table}[!ht]
	\small
	\begin{center}
		\caption{ Comparison between the observational and synthetic
			magnitudes, colors and magnitude differences of the system Hip 12552.} \label{synth3}
		\begin{tabular}{lcc}
			\noalign{\smallskip}
			\hline
			\noalign{\smallskip}
			& Observed $^a$ & Synthetic $^{b}$ (This work) \\
			\hline
			\noalign{\smallskip}
			$V_{J}$ & $8\fm51$ & $8\fm51\pm0.03$\\
			$B_T$  & $9\fm24\pm0.02$   &$9\fm25\pm0.03$\\
			$V_T$  & $8\fm59\pm0.01$   &$8\fm58\pm0.03$\\
			$(B-V)_{J}$&$ 0\fm60\pm0.02$ &$ 0\fm60\pm0.03$\\
			$\triangle m$  &$ 0\fm76^{c}\pm0.03$  &$ 0\fm76\pm0.04$\\
			\hline \noalign{\smallskip}
		\end{tabular}\\
		$^a$ See Table~\ref{tabl2}\\
		$^b$ See Table~\ref{synth1}\\
		$^c$ As the average of all $\Delta m $ using the differnt filters under V-band (see Table~\ref{tabl4}).
	\end{center}
\end{table}
\begin{table}[!ht]
	\small
	\begin{center}
		\caption{Parameters of the components of the system HIP12552.}
		\label{tablef1}
		\begin{tabular}{lcc}
			\noalign{\smallskip}
			\hline\hline
			\noalign{\smallskip}
			Parameters & Comp. a &  Comp. b  \\
			\hline
			\noalign{\smallskip}
			$T_{\rm eff}$\,(K) & $6180\pm50$ & $5865\pm70$ \\
			Radius (R$_{\odot}$) & $1.262\pm0.08$ & $1.006\pm0.07$ \\
			$\log g$ & $4.35\pm0.12$ & $4.45\pm0.14$ \\
			$L (L_\odot)$ & $2.09\pm0.10 $  & $1.08\pm0.12$\\
			$M_{bol}$ & $3\fm95\pm0.18$ & $4\fm67\pm0.19$\\
			$M_{V}$ & $4\fm07\pm0.18$ & $4\fm83\pm0.19$\\
			Mass ($M_{\odot})^{*}$& $1.17 \pm0.11$ & $1.06 \pm0.10$  \\
			Sp. Type$^{**}$ & F8 & G2 \\
			\hline
			\multicolumn{1}{l}{Parallax (mas) }
			& \multicolumn{2}{c}{$11.83\pm 1.07 $}\\

			\multicolumn{1}{l}{($\frac{M_a+M_b }{M_{\odot}}$)$^{***}$ }
			& \multicolumn{2}{c}{$2.23 \pm 0.57 $}\\

			%\multicolumn{1}{l}{Age (Gy) }
			%& \multicolumn{2}{c}{ $2.7\pm 0.3$}\\
			\hline\hline
			\noalign{\smallskip}
		\end{tabular}\\
		$^{*}${depending on the evolutionary tracks (Fig.~\ref{evol})},\\
		$^{**}${depending on the tables of \cite{1992adps.book.....L}},\\
		$^{***}${depending on the orbital solution}.	
	\end{center}
\end{table}	
\section{Results and Discussion}
The synthetic SEDs of the individual component and the system HIP12552 are built using atmospheric modeling and the visual magnitude difference between the two components along with the total observed SED. Least square fitting with weights inversely proportional to the squares of the positional measurement errors is used to modify the orbit of the system. So, the physical and geometrical parameters of HIP12552 are estimated.
Fig.~\ref{fig.} shows the best fit of the total synthetic SED to the observed one.

Table~\ref{synth3} shows a comparison between the observational and synthetic magnitudes, colors and magnitudes differences for the system HIP12552. This gives a good indication of the reliability of the estimated parameters of the individual components of the system which are listed in Table~\ref{tablef1}.

\begin{figure}
	\centering
	%	 \centerline{\psfig{figure=evol.eps,width=0.5\textwidth,clip=}}
	\includegraphics[angle=0,width=14cm]{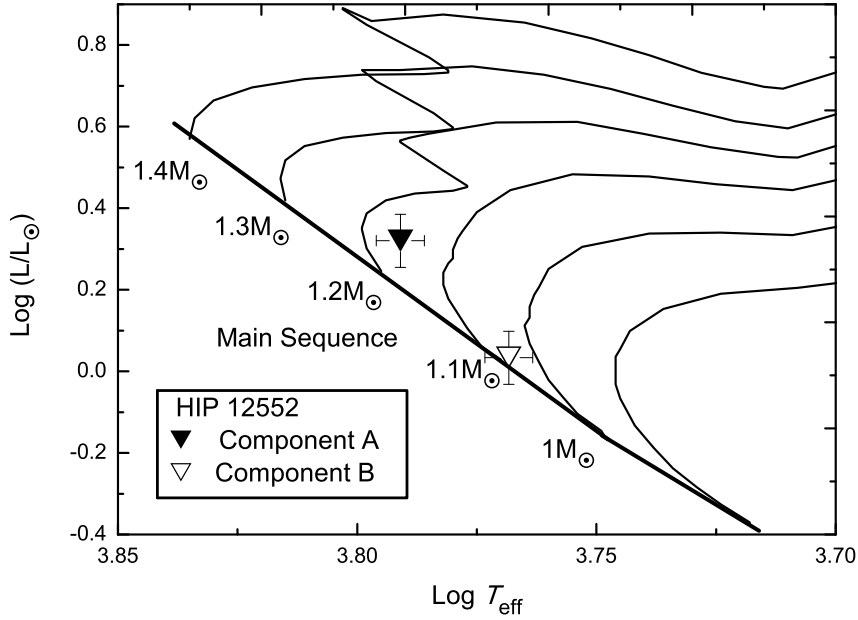}
	\caption{The  system components  on the evolutionary tracks of ~\cite{2000yCat..41410371G}. }
	\label{evol}
\end{figure}

\begin{figure}
	\resizebox{\hsize}{!} {\includegraphics[]{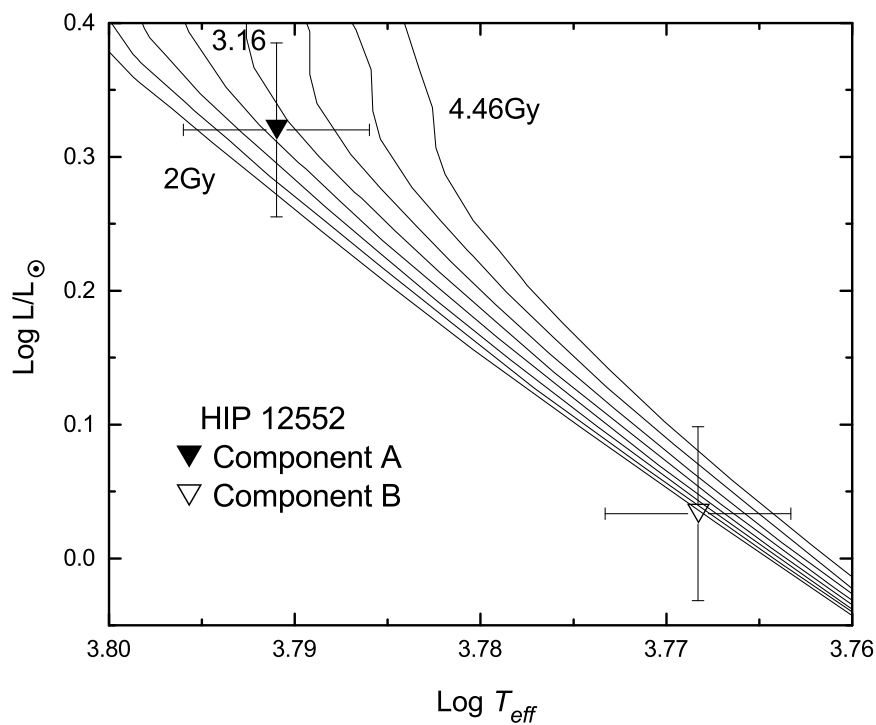}}
	\caption{The systems' components on the isochrones of low- and intermediate-mass, solar composition [$Z$=0.019, $Y$=0.273] stars of ~\cite{2000A&AS..141..371G}.}
	\label{isoch}
\end{figure}

\begin{figure}
	\resizebox{\hsize}{!} {\includegraphics[]{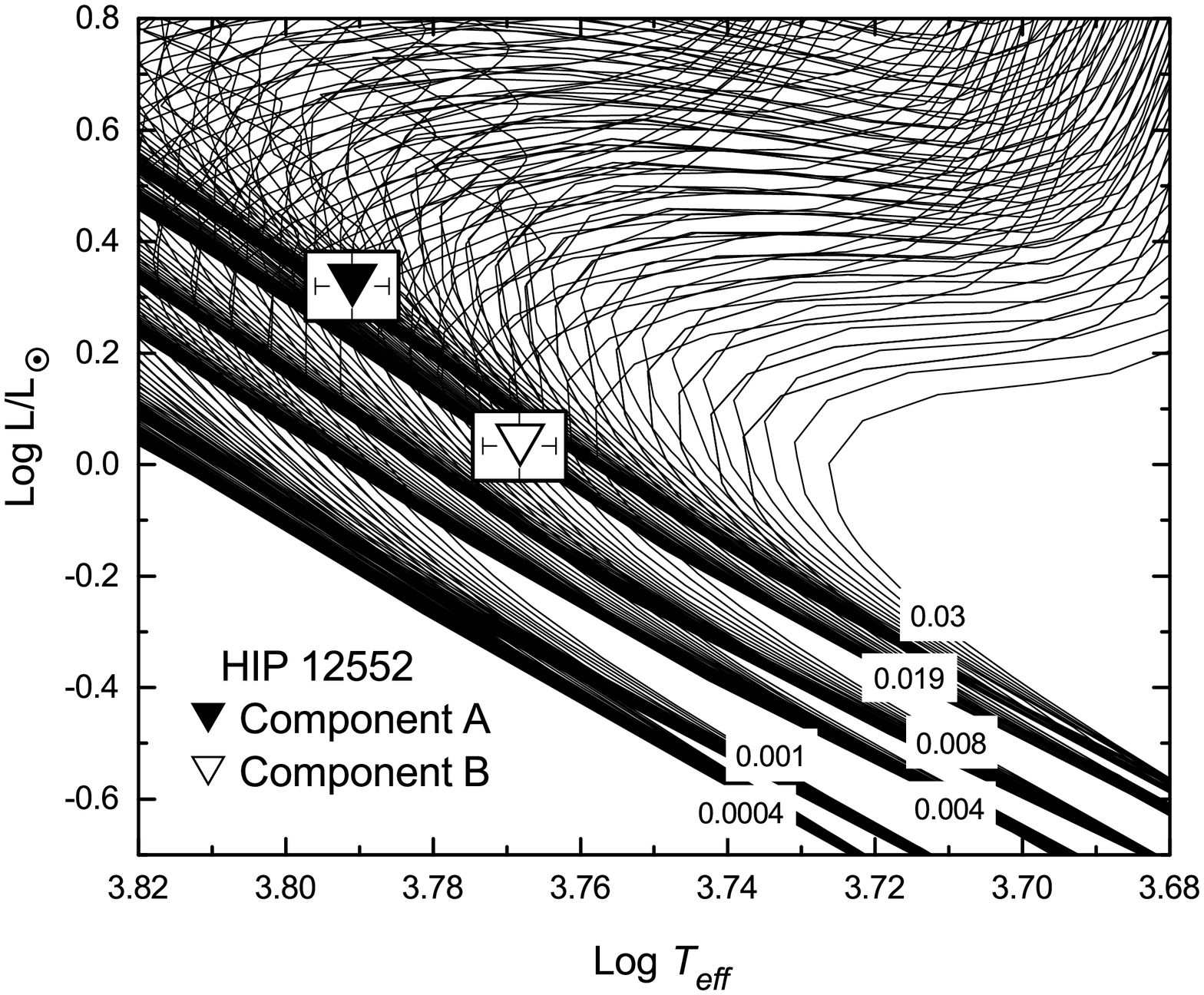}}
	\caption{The systems' components on the isochrones for low- and intermediate-mass stars of different metallicities of  ~\cite{2000A&AS..141..371G}.}
	\label{isoch2}
\end{figure}

The positions of the system's components on the evolutionary tracks of   \cite{2000A&AS..141..371G} (Fig.~\ref{evol}) show that both components with masses  $M_A=1.20$ and $M_B=1.09 M_\odot$ belong to the main-sequence stars. And their positions on   ~\cite{2000A&AS..141..371G} isochrones for low- and intermediate-mass stars of different metallicities and that of the solar composition [$Z=0.019, Y=0.273$] are shown in Figs ~\ref{isoch} \& ~\ref{isoch2}, which give an age of the system around $3.0\pm 0.9$ Gy.

\section{Conclusion}
The CVBS COU1511 (HIP12552) is analyzed  using Al-Wardat's complex method for analyzing close visual binary stars, which is based on combining magnitude difference measurements from speckle interferometry,  entire spectral energy distribution (SED) from spectrophotometry, atmospheres modeling and orbital analysis to estimate the individual physical and geometrical parameters of the system.

 The entire and individual  Johnson-Cousin UBVR,  Str\"{o}mgren uvby, and  Tycho BV synthetic magnitudes and color indices of the system are calculated.  A modified orbit and geometrical elements of the system are introduced and compared with earlier results.

 The positions of the two components on the evolutionary tracks and isochrones are shown, their spectral types  are estimated as F8 and G1 respectively with the age of $3.0\pm 0.9$ Gy.
	
\begin{acknowledgements}
	%\section*{Acknowledgments}
	This work made use of  SAO/ NASA,  SIMBAD database, Fourth Catalog of Interferometric Measurements of Binary Stars, IPAC
	data systems and CHORIZOS code of photometric and spectrophotometric data analysis. The authors thank Mrs. Kawther Al-Waqfi for her help in some orbital calculations.
\end{acknowledgements}

\newpage
%\bibliographystyle{aa}
%\bibliographystyle{raa}
%\bibliography{refernces}

\begin{thebibliography}{32}
\providecommand\natexlab[1]{#1}
\providecommand\JournalTitle[1]{#1}

\bibitem[{Al-Wardat}(2002)]{2002BSAO...53...58A}
{Al-Wardat}, M.~A. 2002, Bull.~Special Astrophys.~Obs., 53, 58

\bibitem[{Al-Wardat}(2012)]{2012PASA...29..523A}
{Al-Wardat}, M.~A. 2012, \pasa, 29, 523

\bibitem[{Al-Wardat} {et~al.}(2014)]{2014AstBu..69...58A}
{Al-Wardat}, M.~A., {Balega}, Y.~Y., {Leushin}, V.~V., {et~al.} 2014,
  Astrophysical Bulletin, 69, 58

\bibitem[{Al-Wardat} \& {Widyan}(2009)]{2009AstBu..64..365A}
{Al-Wardat}, M.~A., \& {Widyan}, H. 2009, Astrophysical Bulletin, 64, 365

\bibitem[{Balega} {et~al.}(2006)]{2006BSAO...59...20B}
{Balega}, I.~I., {Balega}, A.~F., {Maksimov}, E.~V., {et~al.} 2006,
  Bull.~Special Astrophys.~Obs., 59, 20

\bibitem[{Balega} {et~al.}(1994)]{1994A&AS..105..503B}
{Balega}, I.~I., {Balega}, Y.~Y., {Belkin}, I.~N., {et~al.} 1994, \aaps, 105,
  503

\bibitem[{Balega} {et~al.}(2013)]{2013AstBu..68...53B}
{Balega}, I.~I., {Balega}, Y.~Y., {Gasanova}, L.~T., {et~al.} 2013,
  Astrophysical Bulletin, 68, 53

\bibitem[{Balega} {et~al.}(2007)]{2007AstBu..62..339B}
{Balega}, I.~I., {Balega}, Y.~Y., {Maksimov}, A.~F., {et~al.} 2007,
  Astrophysical Bulletin, 62, 339

\bibitem[{Balega} {et~al.}(1999)]{1999A&AS..140..287B}
{Balega}, I.~I., {Balega}, Y.~Y., {Maksimov}, A.~F., {et~al.} 1999, \aaps, 140,
  287

\bibitem[{Couteau}(1996)]{1996IAUDS.128....1C}
{Couteau}, P. 1996, IAU Commission on Double Stars, 128, 1

\bibitem[{Docobo} {et~al.}(2001)]{2001A&A...366..868D}
{Docobo}, J.~A., {Tamazian}, V.~S., {Balega}, Y.~Y., {et~al.} 2001, \aap, 366,
  868

\bibitem[{Docobo} {et~al.}(2006)]{2006AJ....132..994D}
{Docobo}, J.~A., {Tamazian}, V.~S., {Balega}, Y.~Y., \& {Melikian}, N.~D. 2006,
  \aj, 132, 994

\bibitem[{ESA}(1997)]{1997yCat.1239....0E}
{ESA}. 1997, {The Hipparcos and Tycho Catalogues (ESA)}

\bibitem[{Gili} \& {Prieur}(2012)]{2012AN....333..727G}
{Gili}, R., \& {Prieur}, J.-L. 2012, Astronomische Nachrichten, 333, 727

\bibitem[{Girardi} {et~al.}(2000{\natexlab{a}})]{2000A&AS..141..371G}
{Girardi}, L., {Bressan}, A., {Bertelli}, G., \& {Chiosi}, C.
  2000{\natexlab{a}}, \aaps, 141, 371

\bibitem[{Girardi} {et~al.}(2000{\natexlab{b}})]{2000yCat..41410371G}
{Girardi}, L., {Bressan}, A., {Bertelli}, G., \& {Chiosi}, C.
  2000{\natexlab{b}}, VizieR Online Data Catalog, 414, 10371

\bibitem[{Gray}(2005)]{2005oasp.book.....G}
{Gray}, D.~F. 2005, {The Observation and Analysis of Stellar Photospheres}, 505

\bibitem[{Hartkopf} \& {Mason}(2001)]{2001IAUDS.145....1H}
{Hartkopf}, W.~I., \& {Mason}, B.~D. 2001, IAU Commission on Double Stars, 145,
  1

\bibitem[{Hartkopf} {et~al.}(2008)]{2008AJ....135.1334H}
{Hartkopf}, W.~I., {Mason}, B.~D., \& {Rafferty}, T.~J. 2008, \aj, 135, 1334

\bibitem[{Hartkopf} {et~al.}(1994)]{1994AJ....108.2299H}
{Hartkopf}, W.~I., {McAlister}, H.~A., {Mason}, B.~D., {et~al.} 1994, \aj, 108,
  2299

\bibitem[{Hartkopf} {et~al.}(1997)]{1997AJ....114.1639H}
{Hartkopf}, W.~I., {McAlister}, H.~A., {Mason}, B.~D., {et~al.} 1997, \aj, 114,
  1639

\bibitem[{Horch} {et~al.}(2011)]{2011AJ....141...45H}
{Horch}, E.~P., {Gomez}, S.~C., {Sherry}, W.~H., {et~al.} 2011, \aj, 141, 45

\bibitem[{Kurucz}(1994)]{1994KurCD..19.....K}
{Kurucz}, R. 1994, Solar abundance model atmospheres for 0,1,2,4,8 km/s.~Kurucz
  CD-ROM No.~19.~ Cambridge, Mass.: Smithsonian Astrophysical Observatory,
  1994., 19

\bibitem[{Lang}(1992)]{1992adps.book.....L}
{Lang}, K.~R. 1992, {Astrophysical Data I. Planets and Stars.}, 133

\bibitem[{Ma{\'{\i}}z Apell{\'a}niz}(2006)]{2006AJ....131.1184M}
{Ma{\'{\i}}z Apell{\'a}niz}, J. 2006, \aj, 131, 1184

\bibitem[{Ma{\'{\i}}z Apell{\'a}niz}(2007)]{2007ASPC..364..227M}
{Ma{\'{\i}}z Apell{\'a}niz}, J. 2007, in Astronomical Society of the Pacific
  Conference Series, Vol. 364, The Future of Photometric, Spectrophotometric
  and Polarimetric Standardization, ed. C.~{Sterken} (San Francisco:
  Astronomical Society of the Pacific), 227

\bibitem[{Mason} {et~al.}(2011)]{2011AJ....142..176M}
{Mason}, B.~D., {Hartkopf}, W.~I., {Raghavan}, D., {et~al.} 2011, \aj, 142, 176

\bibitem[{McAlister} {et~al.}(1987)]{1987AJ.....93..688M}
{McAlister}, H.~A., {Hartkopf}, W.~I., {Hutter}, D.~J., \& {Franz}, O.~G. 1987,
  \aj, 93, 688

\bibitem[{McAlister} \& {Hendry}(1982)]{1982ApJS...49..267M}
{McAlister}, H.~A., \& {Hendry}, E.~M. 1982, \apjs, 49, 267

\bibitem[{ten Brummelaar} {et~al.}(2000)]{2000AJ....119.2403T}
{ten Brummelaar}, T., {Mason}, B.~D., {McAlister}, H.~A., {et~al.} 2000, \aj,
  119, 2403

\bibitem[{Tokovinin}(1992)]{1992ASPC...32..573T}
{Tokovinin}, A. 1992, in Astronomical Society of the Pacific Conference Series,
  Vol.~32, IAU Colloq. 135: Complementary Approaches to Double and Multiple
  Star Research, ed. H.~A. {McAlister} \& W.~I. {Hartkopf}, 573

\bibitem[{van Leeuwen}(2007)]{2007A&A...474..653V}
{van Leeuwen}, F. 2007, \aap, 474, 653

\end{thebibliography}

\end{document}